\pdfoutput=1
\documentclass[aps,prl,twocolumn]{revtex4-1}
\usepackage{color,amsthm,amsmath,amsfonts,graphicx,bm,epsfig,float,siunitx,multirow,xr,enumerate,epstopdf,soul}
\usepackage[breaklinks=true,colorlinks=true,linkcolor=blue,urlcolor=blue,citecolor=blue]{hyperref}
\usepackage[T1]{fontenc}
\usepackage{physics,verbatim,ulem,mathrsfs}
\begin{document}

\title{Observation of Exceptional Points in Thermal Atomic Ensembles}
\author{Chao Liang$^{1}$}
\author{Yuanjiang Tang$^{1}$}
\author{An-Ning Xu$^{1}$}
\author{Yong-Chun Liu$^{1,2}$}
\email{ycliu@tsinghua.edu.cn}


\affiliation{$^{1}$State Key Laboratory of Low-Dimensional Quantum Physics, Department of
Physics, Tsinghua University, Beijing 100084, China}
\affiliation{$^{2}$Frontier Science Center for Quantum Information, Beijing
100084, China}
\date{\today}

\begin{abstract}
Exceptional points (EPs) in non-Hermitian systems have recently attracted
wide interests and spawned intriguing prospects for enhanced sensing.
However, EPs have not yet been realized in thermal atomic ensembles, which
is one of the most important platforms for quantum sensing. Here we
experimentally observe EPs in multi-level thermal atomic ensembles, and
realize enhanced sensing of magnetic field for one order of magnitude. We take advantage of the rich
energy levels of atoms and construct effective decays for selected energy
levels by employing laser coupling with the excited state, yielding
unbalanced decay rates for different energy levels, which finally results in
the existence of EPs. Furthermore, we propose the optical polarization
rotation measurement scheme to detect the splitting of the resonance peaks,
which makes use of both the absorption and dispersion properties, and shows
advantage with enhanced splitting compared with the conventional
transmission measurement scheme. Besides, in our system both the effective
coupling strength and decay rates are flexibly adjustable, and thus the
position of the EPs are tunable, which expands the measurement range. Our
work not only provides a new controllable platform for studying EPs and
non-Hermitian physics, but also provide new ideas for the design of
EP-enhanced sensors and opens up realistic opportunities for practical
applications in the high-precision sensing of magnetic field and other
physical quantities.
\end{abstract}

\maketitle

Non-Hermitian physics has been one of the research highlights in recent
years \cite{Bergholtz2021,Ashida2020}. Compared with the Hermitian
Hamiltonians, non-Hermitian Hamiltonians have many interesting and unique
properties, where one of the prominent examples is non-Hermitian
degeneracies, also known as exceptional points (EPs) \cite%
{Feng2017,El-Ganainy2018,Miri2019}. An EP occurs when two or more
eigenvalues and the corresponding eigenstates coalesce, simultaneously,
which is impossible for Hermitian Hamiltonians. In the vicinity of EPs,
complex energies of a non-Hermitian system can lead to novel phenomenon
which can not appear in their Hermitian counterparts. For example, when two
degenerate eigenmodes are lifted by a perturbation $\epsilon $, the
eigenfrequency splitting $\Delta \omega $ satisfies a square-root law, i.e.,
$\Delta \omega \propto \sqrt{\epsilon }$, which is very different from
Hermitian cases where signals scale linearly with the perturbation $\epsilon $
\cite{Vollmer2008,Zhu2010,Zijlstra2012,Zhi2017}. Obviously, this sublinear
response signifies an enhanced measurement sensitivity $\propto 1/\sqrt{%
\epsilon }$ in the small perturbation limit $\epsilon \rightarrow 0$, which
can be used to design EP-enhanced sensors. In recent years, EPs have been
studied in many systems, e.g., optical microcavities \cite%
{PhysRevLett.112.203901,Lai2019,Wang2020i,zhang2018phonon,Lee2009,Chen2017,Hodaei2017,Wang2021,Peng2014a,Chang2014}%
, photonic crystal slabs \cite{Zhen2015,Park2020}, acoustic systems \cite%
{Ding2016}, circuit \cite%
{Kononchuk2022,Yang2022,Xiao2019a,Sun2014,Assawaworrarit2017,Chen2018a},
optomechanical systems \cite{PhysRevLett.113.053604,Xu2016a}, superconducting systems \cite%
{Naghiloo2019,Chen2022,Abbasi2021a}, ultracold atoms \cite{Li2019}, trapped
ions \cite{Ding2021}, atomic systems \cite{Lefebvre2009,Cartarius2007,Magunov1999,Latinne1995,Berry1998,Peng2016a} and nitrogen-vacancy centers \cite{Wu2019b}.

However, EPs have not yet been realized in thermal atomic ensembles, which
is one of the most important platforms for exploring quantum precision
measurement and quantum sensors, e.g., ultrasensitive magnetometer \cite%
{Kominis2003,Budker2007,Bao2020}, gyroscope \cite{wu2020}, electrometer \cite%
{Jing2020a,Sedlacek2012b} and atomic clock \cite{Cutler_2005}. 
Therefore, it is urgent to achieve EPs and design
EP-enhanced sensing schemes in thermal atomic ensembles, so that we can take
advantage of the non-Hermitian features for practical applications in a vast
variety of sensors.



Here we propose a new paradigm for studying the non-Hermitian physics by
taking advantage of the rich energy level structure and couplings in thermal
atomic ensembles. We experimentally observe EPs in thermal atomic ensembles
and realize enhanced sensing of magnetic field, for the first time as far as
we know. Moreover, instead of measuring transmission spectrum in
conventional studies, we propose a new protocol relied on the optical
polarization rotation (OPR) signal to detect the resonance peaks splitting, which
can enhance the frequency splitting and is robustness to the noises. We
demonstrate that peak splitting of the optical rotation signal scales as the
square root of the perturbation magnetic-field strength and maintain a high
sensitivity for weak perturbation. In addition, in our system most
experimental parameters are flexibly adjustable, so we can move the position
of the EP to expand the measurement range of magnetic field. Therefore, this
work opens up realistic opportunities for practical applications in
high-precision sensing of magnetic field.

\begin{figure}[tb]
\centering
\includegraphics[width=\linewidth]{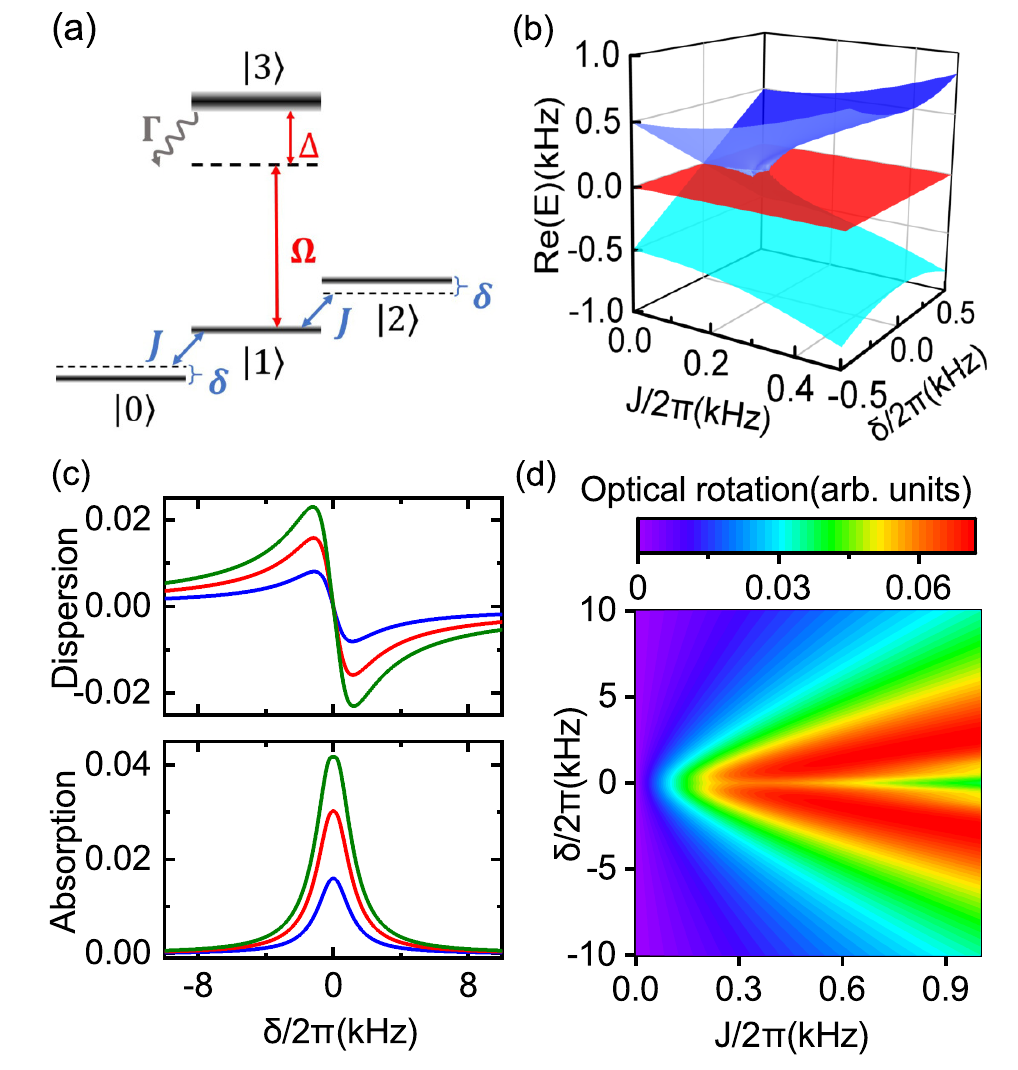}
\caption{System model for observing EPs in atomic ensembles. (a) Atomic
energy level diagrams. $\Gamma $ denotes the spontaneous decay rate of the
excited state $|3\rangle $. $J$ denotes the effective coupling rate from an
applied oscillating RF magnetic field with frequency detuning $\protect%
\delta $ relative to the transition of adjacent Zeeman levels $|0\rangle $, $%
|1\rangle $, $|2\rangle $. The probe lasers drive the transition $|1\rangle
\Leftrightarrow |3\rangle $ with the detuning $\Delta $ and the effective
Rabi frequencies is $\Omega $. (b) The real parts of the eigenvalues of $H_{%
\mathrm{NH}}$ in Eq. (\protect\ref{eq1}) as a function of $\protect\delta $
and $J$. (c) The absorption $S_{\mathrm{abs}}$ and dispersion $S_{\mathrm{disp}}$ curves of the optical
polarization rotation signal for different RF Rabi frequency $J/2\protect\pi
$ being $0.05$ \textrm{kHz} (blue curve), $0.10$ \textrm{kHz }(red curve)
and $0.15$ \textrm{kHz }(green curve), respectively. (d) Color map of the magnitude of the optical polarization rotation signal versus detuning $\protect\delta $ for various values of $J$. Other parameters are specified
in the main text.}
\label{fig1}
\end{figure}

Our model is based on a four-level atomic system as shown in Figs.~\ref{fig1}%
(a). There are three Zeeman sublevels of ground state $|0\rangle $, $%
|1\rangle $, and $|2\rangle $, which are coupled by an oscillating
radio-frequency (RF) magnetic field. A laser drives the transition from
ground state $|1\rangle $ to the excited state $|3\rangle $. Their Rabi
frequencies and the detunings of coupling fields above from their coupled
transitions are denoted by $(J_{0},J_{0},\Omega _{0})$ and $(\delta ,\delta
,\Delta )$, respectively. The spontaneous decay rate of the excited state $%
|3\rangle $ is $\Gamma $ and the relaxation rate of the Zeeman sublevels is $%
\gamma _{0}$, which satisfies $\gamma _{0}\ll \Gamma $. In the rotating
reference frame, the Hamiltonian reads \cite{supp}: 
\begin{equation}
H=\left(
\begin{array}{cccc}
-\delta & J & 0 & 0 \\
J & 0 & J & -\Omega \\
0 & J & \delta & 0 \\
0 & -\Omega & 0 & -\Delta%
\end{array}%
\right) .
\end{equation}%
where $J=J_{0}/(2\sqrt{2})$ is the effective RF Rabi frequencies and $\Omega
=\Omega _{0}/(2\sqrt{3})$ is the effective optical Rabi frequencies. The
evolution of the density matrix $\rho $ is described by the master equation $%
\dot{\rho}=-i[H,\rho ]+\mathcal{L}[\rho ]$, where $\mathcal{L}$ is Lindblad
operator describing the decay and dephasing of the system. When the decay
rate $\Gamma $ is much greater than all other rates, we can eliminate the
excited state $|3\rangle $ and obtain the effective non-Hermitian
Hamiltonian $H_{\mathrm{NH}}$ governing the dynamics of the ground sublevels
\cite{supp} (set $\Delta =0$ for simplicity)
\begin{equation}
H_{\mathrm{NH}}=\left(
\begin{array}{ccc}
-\delta & J & 0 \\
J & -\frac{i}{2}\gamma_\mathrm{opt} & J \\
0 & J & \delta%
\end{array}%
\right) .  \label{eq1}
\end{equation}%
Here $\gamma_\mathrm{opt} =4\Omega ^{2}/\Gamma \ $corresponds to an effective decay rate
for state $|1\rangle $, which is the key factor to realize the effective
non-Hermitian Hamiltonian. 
This effective decay rate $\gamma_\mathrm{opt}$ originates from the
laser-driven coupling between state $|1\rangle $ and state $|3\rangle $,
where the strong decay of state $|3\rangle $ results in the effective decay
of state $|1\rangle $. Similar effective non-Hermitian Hamiltonian have been studied in the anti-parity-time symmetry system \cite{Zhang2019,PhysRevA.96.053845,PhysRevLett.129.273601}.

When $\delta =0$, the three eigenvalues of $H_{\mathrm{NH}}$ are given by
\begin{equation}
E_{0}=0,\quad E_{\pm }=-i\kappa \gamma _{0}\pm \sqrt{2J^{2}-\kappa
^{2}\gamma _{0}^{2}},  \label{eq2}
\end{equation}%
where $\kappa =\Omega ^{2}/(\Gamma \gamma _{0})$ is dimensionless saturation
parameter of the probe laser. As shown in Eq.~\ref%
{eq2}, in the case of $J=J_{\mathrm{EP}}\equiv \kappa \gamma _{0}/\sqrt{2}$,
the real (Fig.~\ref{fig1}(b)) and imaginary parts of the eigenvalues $E_{\pm }$ will degenerate
simultaneously, corresponding to the EPs. Thus the system possesses
second-order EPs and it behaves like a $\mathcal{PT}$-symmetry two-level
system \cite{Mandal2021a}. When $J<J_{\mathrm{EP}}$, the complex eigenvalues
have different imaginary parts, and the system is in the $\mathcal{PT}$%
-symmetry-broken-like phase. On crossing the EP with $J>J_{\mathrm{EP}}$, the
imaginary parts of the eigenvalues $E_{\pm }$ coincide, and the system is in
the $\mathcal{PT}$-symmetry-like phase.

In the parametric space consisting of the effective RF Rabi frequencies $J$
and the saturation parameters $\kappa $, the EPs are joined to form
exceptional arcs which satisfy $J=\kappa \gamma _{0}/\sqrt{2}$. Remarkably, in our system the parameters $J$ and $\kappa
$ can be freely tuned by adjusting the RF and optical driving strength,
which correspond to the magnetic filed strength and laser power,
respectively. Thus it enables free control of the position of EPs, and is
promising for designing EP-enhanced sensors with broad measurement range.

Furthermore, third-order EPs can also be obtained as long as the sublevels
of ground state $|0\rangle $, $|1\rangle $, and $|2\rangle $ have different
effective decay rates, which can be realized by using similar laser-driven
couplings $|0\rangle $ $\Leftrightarrow $ $|3\rangle $ and $|2\rangle $ $%
\Leftrightarrow $ $|3\rangle $ \cite{supp}. This provides more possibilities
for studying high-order non-Hermitian physics with thermal atomic systems.

The system eigenstates can be experimentally probed by slowly sweeping the
detuning $\delta $ of the RF magnetic field with time and measuring the optical
polarization rotation (OPR) signal of the probe laser. The magnitude of the
OPR signal is composed of absorptive part and
dispersive part. The expectation value of absorptive ($S_{\mathrm{abs}}$)
and dispersive signal ($S_{\mathrm{dis}}$) is found from the density matrix
of ground states \cite{supp}:
\begin{equation}
\begin{aligned} S_{\mathrm{abs}}&\propto \mathrm{Re}(\rho_{10}-\rho_{21}),\\
S_{\mathrm{dis}}&\propto \mathrm{Im}(\rho_{10}-\rho_{21}), \end{aligned}
\label{S}
\end{equation}%
where density matrix elements $\rho _{10}$ and $\rho _{21}$ represent the
coherences between Zeeman sublevels. Typical absorption and dispersion
profiles versus $\delta $ are shown in Fig.~\ref{fig1}(c). We can further
define the magnitude of the OPR signal as $S=\sqrt{%
S_{\mathrm{abs}}^{2}+S_{\mathrm{dis}}^{2}}$. The color map of optical
polarization rotation magnitude versus $J$ and $\delta $ is plotted in Fig.~%
\ref{fig1}(d). It shows that as $J$ increases, the resonance peak splits
into two peaks, and the positions of the peaks can be derived as \cite{supp}%
:
\begin{equation}
f_{\pm }(J)=%
\begin{cases}
0 & \mbox{if}\quad J<J_{\mathrm{OPR}}, \\
\pm \sqrt{2J_{\mathrm{OPR}}\left( J-J_{\mathrm{OP\mathrm{R}}}\right) } & %
\mbox{if}\quad J\geq J_{\mathrm{OPR}},%
\end{cases}%
\end{equation}%
where $J_{\mathrm{OPR}}$ is the peak splitting parameter of the OPR signal with the expression $J_{\mathrm{OPR}}=\gamma _{0}\sqrt{(8\kappa ^{2}+6\kappa +1)/(8\kappa
(16\kappa (2\kappa +3)+21)+26)}$. It reveals that the peak positions $f_{\pm }(J)
$ have square-root response when $J\geq J_{\mathrm{OPR}}$, and such a
sublinear response provides the opportunity for EP-enhanced sensing.

\begin{figure}[tb]
\centering
\includegraphics[width=\linewidth]{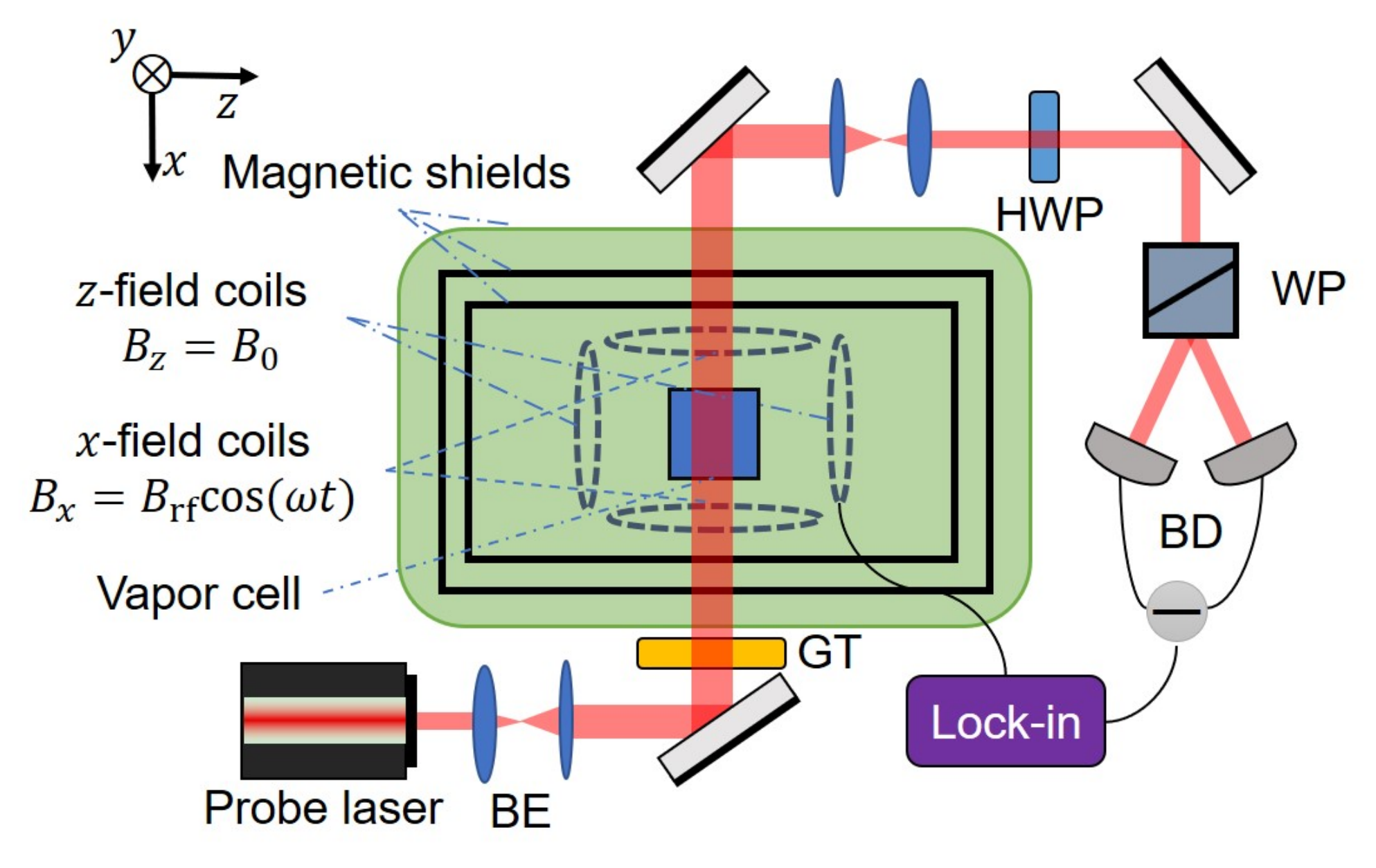}
\caption{Schematic diagram of the experimental setup. A laser beam linearly
polarized along the $z$ axis propagates through a paraffin-coated vapor cell
which is filled with rubidium-87 ($^{87}$Rb) atoms. A constant $z$-direction
magnetic field $B_{0}$ and an oscillating $x$-direction RF magnetic field $%
B_{x}=B_{\mathrm{rf}}\cos (\protect\omega t)$ are applied within the
magnetic shields which surround the cell. The lock-in amplifier is used to
analyze the OPR signal. BE: beam expansion module;
GT: Glan-Taylor polarizer; HWP: half wave plate; WP: Wollaston prism; BD:
balanced detection.}
\label{fig2}
\end{figure}

Based on the above theoretical model, we design the experiment with the
schematic diagram shown in Fig.~\ref{fig2}. A cubic paraffin-coated glass
cell (side length 1 cm), surrounded by a two-layer $\mu $-metal magnetic
shield, is filled with rubidium-87 ($^{87}$Rb) atoms. The temperature of the
cell is stabilized at 50$^{\circ }$C. The combination of one pair of
Helmholtz coil and two pairs of gradient and uniform saddle coils \cite%
{Jeon2010}, which is placed inside the innermost shield, are used to
compensate the residual magnetic fields and to create static ($B_{z}=B_{0}$
in the $z$ direction) and oscillating fields ($B_{x}=B_{\mathrm{rf}}\cos
(\omega t)$ in the $x$ direction). The typical static fields strength $B_{0}$
is $0.65$ $\mathrm{G}$, which gives a ground-state Zeeman splitting of $%
\Omega _{L}/2\pi =453$ $\mathrm{kHz}$. A $z-$polarized probe laser
propagating through the vapor cell along the $x-$axis is red-detuned by 700
\textrm{MHz} from $|5^{2}S_{1/2},F=2,m_{F}=0\rangle \Leftrightarrow
|5^{2}P_{1/2},F=1,m_{F}=0\rangle \left( |1\rangle \Leftrightarrow |3\rangle
\right) $ transition of the D1 line. The light intensity is 5 $\mathrm{\mu W}
$ unless otherwise specified and the beam diameter is $\sim 4$ $\mathrm{mm}$%
. After passing through the vapor cell, the polarization of the laser is
analyzed using a balanced polarimeter setup and extracted by the lock-in
amplifier.

\begin{figure}[tb]
\centering
\includegraphics[width=\linewidth]{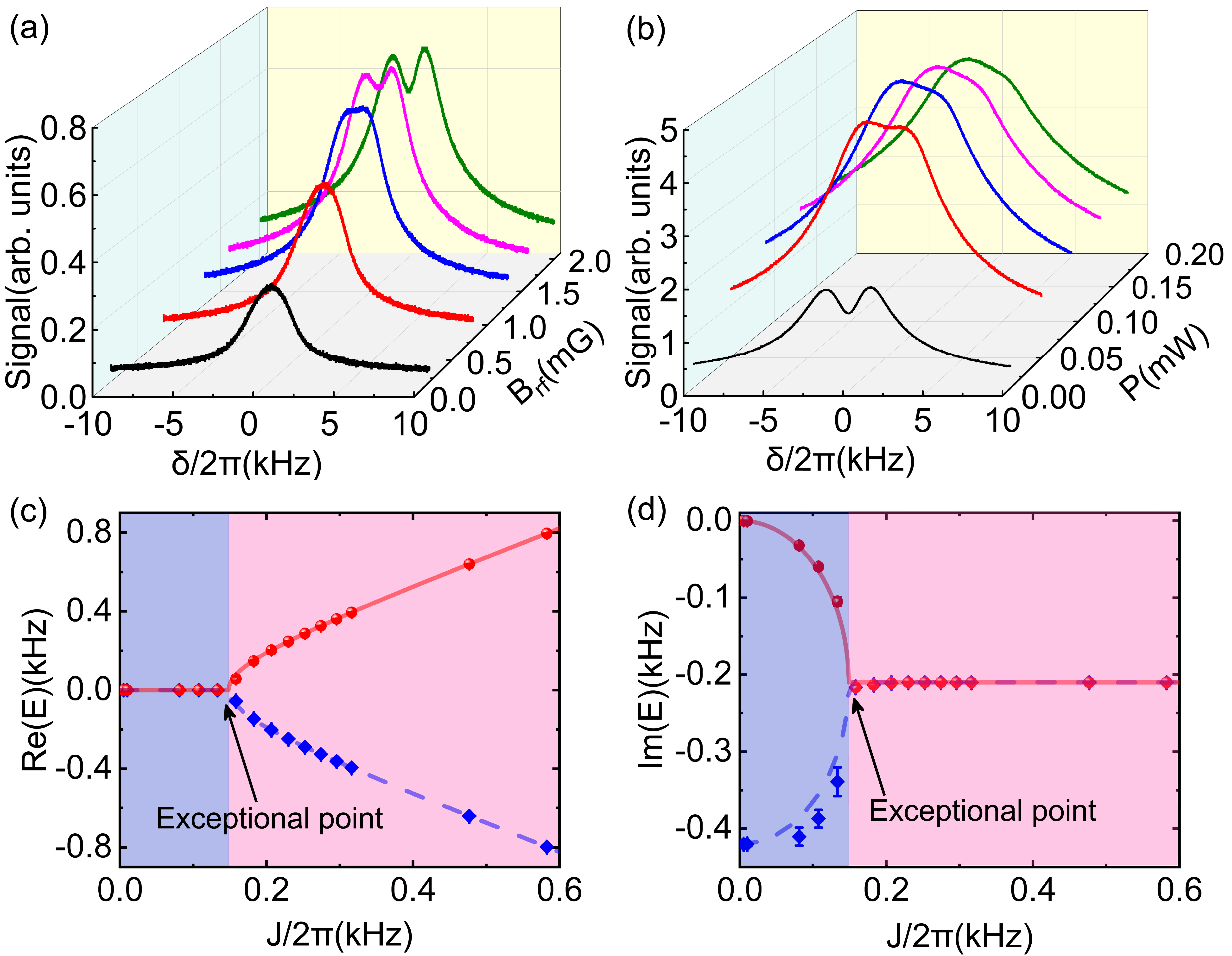}
\caption{Experimental results of EPs. (a)-(b): Experimental results of
OPR signal versus detuning for different RF
magnetic field $B_{\mathrm{rf}}$ from $0.1$ to $2$ \textrm{mG }(a)\textrm{%
, }and for different probe laser power $P$ from $0.005$ to $0.20$ \textrm{mW}
(b). (c)-(d): Experimentally obtained real (c) and imaginary (d) parts of
the system eigenvalues as a function of RF Rabi frequency $J/2\protect\pi $.
The red dots and blue squares are obtained from curve fitting the measured
OPR spectra to the theoretical result. The error
bars are standard deviations obtained from five measurements. The red and
blue curves are obtained from the theoretical model using the experimental
parameters, and we have used $\protect\gamma _{0}/2\protect\pi =0.7$ \textrm{%
kHz}, $\protect\kappa =0.3$. The response curve in (a) is the spectrum corresponding to five typical points in (c)(d). Complete experimental data can be referred to Supplemental Material\cite{supp}}.
\label{fig3}
\end{figure}

\begin{figure*}[tb]
\centering
\includegraphics[width=\linewidth]{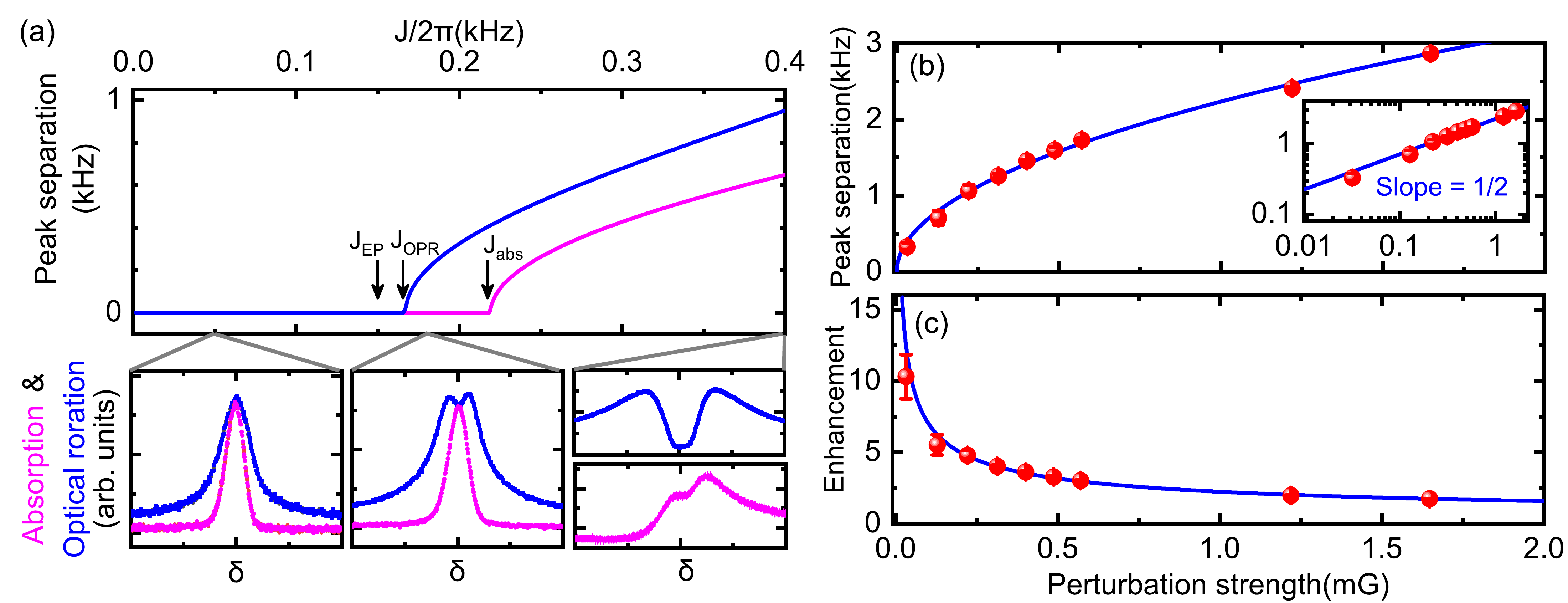}
\caption{EP-Enhanced sensing with OPR measurement.
(a) The peak separation of OPR signal (blue curve)
and absorption spectra (pink curve) as a function of the coupling strength $%
J/2\protect\pi $. Below: snapshots showing the observed optical polarization
rotation and absorption spectra versus the detuning $\protect\delta /2%
\protect\pi $ for three typical coupling strengths, where $J/2\protect\pi$=0.05, 0.18, 0.4 kHz from left to right. (b) The observed peak
separation of OPR magnitude as a function of the
perturbation strength. The inset demonstrates a slope of 1/2 on a
logarithmic scale, confirming the existence of exceptional points. (c) Measured enhancement as a function of the perturbation. The sensitivity demonstrates an order enhancement in the proximity of the EP as opposed to a system configuration away from the EP.
In (b) and (c), the red
dots indicate experimental data and the error bars are standard deviations
obtained from five measurements; the solid blue curves are theoretical
results.
}
\label{fig4}
\end{figure*}

The EPs are observed experimentally using the above setup. In Fig.~\ref{fig3}
(a) we display the measured OPR signal versus
detuning $\delta $ at different RF magnetic field strengths $B_{\mathrm{rf}}$.
As we fix the probe laser power $P$ and increase the strength of the RF
magnetic field $B_{\mathrm{rf}}$, the resonance peak begins to split into
two peaks, which means that we have swept over the EP. On the other hand, as
shown in Fig.~\ref{fig3}(b), the separation distance of the two splitting
peaks will gradually decrease to disappear as we fix the $B_{\mathrm{rf}}$
and increase the probe laser power $P$. To determine the location of the EP,
we display the real (Fig.~\ref{fig3}(c)) and imaginary (Fig.~\ref{fig3}(d)) parts of the eigenvalues $E_{\pm }$
(Eq.~(\ref{eq2})) as a function of $J$ using experimentally obtained values
of saturation parameter $\kappa $ and ground state relaxation rate $\gamma
_{0}$ \cite{Peng2014a,supp}. The blue (pink) shaded regions correspond to
the $\mathcal{PT}$-symmetry-broken-like ($\mathcal{PT}$-symmetry-like) regions and
the boundary represents EP. For typical parameters as used in Fig.~\ref{fig3}(c)-(d),
we obtain $J_{\mathrm{EP}}/2\pi =0.15$ $\mathrm{kHz}$. When $J<J_{%
\mathrm{EP}}$, the two eigenmodes have the same resonance frequencies but
different linewidths. On the other hand, at stronger coupling $J>J_{\mathrm{%
EP}}$, the resonant frequencies of two eigenmodes move in the opposite
direction but their linewidths coincide. We can see that the experiment
results are in good agreement with the theoretical prediction.

The observed EPs in atomic ensembles hold great potential for designing
EP-enhanced magnetic field sensors, as the coupling strength $J$ is directly
related to the magnetic field strength. At the second-order EPs, the
square-root singularity promises greater signal enhancement for a small
perturbation. However, signal enhancement does not always mean increased
sensitivity, and some arguments suggest that the sensitivity of the EP
sensors are degraded by noise \cite{Mortensen2018,Langbein2018,Zhang2019a,Chen2019a,Mortensen2018,Lau2018a}. The
effect of noise on the signal is that the resonance linewidth (imaginary
part of the eigenvalues) increases so that the peak separation is hardly
detected in the experiment even if the resonance frequency (real part of the
eigenvalues) is split. Therefore, eigenfrequency splitting and measured peak
splitting are different. In conventional measurement schemes, the
transmission spectra are used to determine the peak splitting \cite%
{Chen2017,Hodaei2017}, which rely on the transmission peak degeneracies
(TPDs) observed in the transmission spectrum to enhance the signal-to-noise
ratio (SNR) \cite{Kononchuk2022,Geng:21}.

In our scheme, as analyzed in Eq.(\ref{S}), we use the absorption and
dispersion profiles of the OPR signal, and make
use of the OPR peak degeneracies (OPR-PDs) for
observing the peak splitting \cite{supp}. As both absorption and dispersion
properties are considered, the OPR-PDs here is of great advantage for
improving the measurement sensitivity. On one hand, as shown in Fig~\ref{fig4}%
(a), compared with the peak splitting parameter of the absorption signal as $%
J_{\mathrm{abs}}/2\pi =0.22$ $\mathrm{kHz}$ (pink curve), we find that the
peak splitting parameter of OPR signal $J_{\mathrm{%
OPR}}/2\pi =0.17$ $\mathrm{kHz}$ (blue curve) is much closer to the EP
parameter with $J_{\mathrm{EP}}/2\pi =0.15$ $\mathrm{kHz}$, which means that
we obtain a larger signal for the same perturbation. Especially, when $J_{%
\mathrm{OPR}}<J<J_{\mathrm{abs}}$, it clearly shows that the optical
polarization rotation signal is split obviously, while the absorption signal
does not (middle inset of Fig.~\ref{fig4}(a)). On the other hand, since $J_{%
\mathrm{OPR}}$ is slightly larger than $J_{\mathrm{EP}}$, it also avoids the
collapse of eigenbasis at EPs and thus avoids the excess fundamental noise
\cite{Wiersig2020,Wang2020}.

Figure~\ref{fig4}(b) clearly demonstrates a square-root peak splitting in
response to changes in the perturbation magnetic field near $J=J_{\mathrm{OPR%
}}$. As depicted in the inset of Fig.~\ref{fig4}(b), the slope of $1/2$ in the
corresponding logarithmic plot affirms this behaviour.
Thanks to the square-root scaling, enhancement of measurement sensitivity can be realized compared with the conventional linear scaling.
In our experiments, we have observed an order enhancement in the proximity of the EP compared with the linear scaling away from EP, which corresponds to the case when $J\gg J_{\mathrm{OPR}}$ and is the same as the Hermitian case.
The experiment
results are in accordance with the theoretical expectations.

It should be stressed that in our system the parameter $\kappa =\Omega
^{2}/(\Gamma \gamma _{0})$ can be tuned by adjusting the laser power.
Therefore, the corresponding working point $J_{\mathrm{OPR}}$ is real-time
controllable, which largely expands the measurement range and opens up
realistic opportunities for practical applications in absolute magnetic
field measurement in geomagnetism condition.

In summary, we propose a method for studying non-Hermitian physics in
thermal atomic ensembles by making use of the rich energy level structure
and couplings. We experimentally observe EPs and realized enhanced sensing
of magnetic field for one order of magnitude near the EP. We also propose to measure the optical
rotation signal spectrum instead of the transmission spectrum, and
demonstrated that the peak splitting has a square-root dependence on the
perturbation strength, with enhanced sensitivity for detecting magnetic
field. In addition, the position of the EPs can be tuned by adjusting the
laser-driven coupling strength, which expands the measurement range. Our scheme can
also be generalized to realize high-order EPs by introducing more laser-driven couplings.
With the development of advanced nanofabrication technologies, using hollow-core
photonic crystal fibers \cite{PhysRevLett.103.043602,Sprague2014} or atomic
cladding waveguides on a chip \cite{Stern2017,Ritter2018}, our scheme is
also promising for miniaturization and integration. Our work not only
provides a new controllable platform for studying EPs and non-Hermitian
physics, but also opens up an avenue to design high-sensitivity
magnetometers as well as improve the existing measurement methods \cite%
{Jing2020a,Sedlacek2012b}.

\begin{acknowledgments}
This work is supported by the Key-Area Research and Development Program of
Guangdong Province (Grant No.~2019B030330001), the National Natural Science
Foundation of China (NSFC) (Grant Nos. 12275145, 92050110, 91736106, 11674390, and
91836302), and the National Key R\&D Program of China (Grants No.
2018YFA0306504).
\end{acknowledgments}

\bibliography{MainBib}

\end{document}